\documentclass[12pt,british,english]{article}
\usepackage[T1]{fontenc}
\usepackage[latin9]{inputenc}
\usepackage{geometry}
\usepackage{amsmath}
\usepackage{amssymb}

\makeatletter

\newcommand{\noun}[1]{\textsc{#1}}
\providecommand{\tabularnewline}{\\}

\newcommand{\lyxaddress}[1]{
\par {\raggedright #1
\vspace{1.4em}
\noindent\par}
}

\@ifundefined{date}{}{\date{}}
\makeatother

\usepackage{babel}
\begin{document}

\title{Higher Symmetries of the Schrödinger Operator in Newton-Cartan Geometry}

\author{James Gundry}
\maketitle
\begin{abstract}
We establish several relationships between the non-relativistic conformal
symmetries of Newton-Cartan geometry and the Schrödinger equation.
In particular we discuss the algebra $\mathfrak{sch}(d)$ of vector
fields conformally-preserving a flat Newton-Cartan spacetime, and
we prove that its curved \foreignlanguage{british}{generalisation}
generates the symmetry group of the covariant Schrödinger equation
coupled to a Newtonian potential and generalised Coriolis force. We
provide intrinsic Newton-Cartan definitions of Killing tensors and
conformal Schrödinger-Killing tensors, and we discuss their respective
links to conserved quantities and to the higher symmetries of the
Schrödinger equation. Finally we consider the role of conformal symmetries
in Newtonian twistor theory, where the infinite-dimensional algebra
of holomorphic vector fields on twistor space corresponds to the symmetry
algebra $\mathfrak{cnc}(3)$ on the Newton-Cartan spacetime.

\vfill\pagenumbering{gobble} 
\end{abstract}
\noindent \begin{center}
\noun{\small{JAMES.GUNDRY@DAMTP.CAM.AC.UK}}
\par\end{center}{\small \par}

\lyxaddress{\noindent \begin{center}
\noun{\small{Department of Applied Mathematics and Theoretical Physics,
University of Cambridge, Wilberforce Road, Cambridge CB3 0WA, UK.}}
\par\end{center}}

\noindent \begin{center}
\noun{\scriptsize{Reference: DAMTP-2016-2}}
\par\end{center}{\scriptsize \par}

\pagebreak{}

\section{Introduction and Theorems}

\pagenumbering{arabic}One of the chief concerns of mathematical physics
is to put the ubiquitous symmetries of nature on a geometrical footing.
On a Riemannian manifold $(M,g)$, the mathematical setting for relativistic
physics, the central objects in the study of symmetries are Killing
vectors and Killing tensors. The former capture our intuition of what
is meant by a continuous symmetry: if some \foreignlanguage{british}{continuous}
transformation leave a system unchanged then that transformation is
a symmetry of the system. Geometrically this transformation is implemented
by pushing-forward the metric $g$ along the integral curves of a
vector field, i.e. $X$ is a \emph{Killing vector }iff
\begin{equation}
\left(\mathcal{L}_{X}g\right)_{a_{1}a_{2}}=0\,\,,\label{Killingvector}
\end{equation}
where ${\cal L}_{X}$ is the Lie derivative along $X$. The intuitive
concepts of translational and rotational invariance, for instance,
find geometrical guises as Killing vectors. In accordance with Noether's
theorem, Killing vectors usefully correspond to conserved quantities
of the motion.

More exotic are so-called ``hidden'' symmetries. Unlike with Killing
vectors there is no intuitive picture of the metric $g$ being pushed-forward
along integral curves on $M$; instead we must extend our viewpoint
to include the cotangent bundle $T^{*}M$. A hidden symmetry is a
rank-$n$ symmetric contravariant tensor field $X$ which preserves
the metric in the sense that
\begin{equation}
\left\{ X^{a_{1}...a_{n}}p_{a_{1}}...p_{a_{n}}\,,\, g^{b_{1}b_{2}}p_{b_{1}}p_{b_{2}}\right\} =0\,,\label{KillingTensor}
\end{equation}
where $(x^{a},p_{b})\in T^{*}M$ and $\left\{ \,,\,\right\} $ is
the canonical Poisson structure on the cotangent bundle. Whilst the
\emph{Killing tensor} $X$ does not generate transformations via integral
curves on $M$, equation (\ref{KillingTensor}) exploits the fact
that the complete lift
\[
\hat{X}=nX^{a_{1}...a_{n-1}b}p_{a_{1}}...p_{a_{n-1}}\frac{\partial}{\partial x^{b}}-\frac{\partial X}{\partial x^{b}}^{a_{1}...a_{n}}p_{a_{1}}...p_{a_{n}}\frac{\partial}{\partial p_{b}}
\]
of $X$ does generate transformations on $T^{*}M$ \cite{YanoPatterson}.
The associated conserved quantity is the Hamiltonian $X^{a_{1}...a_{n}}p_{a_{1}}...p_{a_{n}}$,
and the existence of such conserved quantities is often essential
to the integrability of geodesic motion. For instance, the Kerr metric
describing a spinning black hole admits a rank-two Killing tensor,
and the associated conserved quantity ``Carter's constant'' allows
one to determine the orbits (see e.g. \cite{Cariglia}). Equation
(\ref{KillingTensor}) can be concisely restated in the language of
the Schouten bracket as $\mathcal{L}_{X}g=0$.

Also of interest are the conformal cousins of Killing vectors and
Killing tensors, arising when we relax equations (\ref{Killingvector}-\ref{KillingTensor})
and allow $X$ to generate conformal transformations of the metric.
The defining condition on the vector $X$ to be a \emph{conformal
Killing vector }or the tensor $X$ to be a \emph{conformal Killing
tensor} then becomes
\[
\mathcal{L}_{X}g^{a_{1}...a_{n+1}}=k^{(a_{1}...a_{n-1}}g^{a_{n}a_{n+1})}
\]
where respectively $\mathcal{L}$ refers to the Lie bracket or the
Schouten bracket, and $k$ is a tensor to be determined. The associated
``conserved quantities'' are conserved only necessarily along null
geodesics. Conformal Killing tensors have appeared in many places
in mathematical physics; for example, as will be relevant in this
paper, conformal Killing tensors generate the higher symmetry operators
of the Laplacian \cite{Eastwood}.

Thus far we have only mentioned Riemannian geometry; the chief aim
of this paper is to build upon the work of \cite{DuvalHovarthy,Son,GibbonsCarigliaetal}
in extending the above ideas to the field of Newton-Cartan geometry.
In particular, a contribution of this paper will be to establish some
intrinsic definitions of Killing tensors and ``conformal'' Schrödinger-Killing
tensors using a Newton-Cartan Hamiltonian formalism and prove two
theorems relating the resulting tensors to the symmetries of the Schrödinger
equation. This Hamiltonian formalism will be in agreement with the
Eisenhart-Duval lift to a Bargmann structure, and the Killing tensors
defined will be in accord with the well-known study of hidden symmetries
in non-relativistic physics, but will be written in the Newton-Cartan
language.

Newton-Cartan geometry is what results when one takes a non-relativistic
limit of general relativity and is the mathematical formalism behind
non-relativistic gravitation \cite{Cartan}. This kind of geometry,
mathematically remarkable because unlike in Riemannian geometry the
connection is non-metric, is of interest to condensed matter physicists
whose theories are non-relativistic. Furthermore Newton-Cartan geometry
has attracted recent attention in attempts to establish a non-relativistic
version of the AdS/CFT correspondence \cite{Son,DuvalLazzarini,Bergshoeff}.
A self-contained introduction to Newton-Cartan spacetimes will be
provided in section two.

A rank-$n$ symmetry ${\cal D}$ of a linear differential operator
$\Delta$ is a linear differential operator of order $n$ which obeys
\begin{equation}
\Delta{\cal D}=\delta\Delta\label{eq:symdeff}
\end{equation}
for some (otherwise irrelevant) linear differential operator $\delta$.
We will be concerned with relating the geometrical non-relativistic
symmetries (such as Schrödinger-Killing tensors) to symmetries in
the sense of (\ref{eq:symdeff}).

In section three we will discuss the well-known Schrödinger algebra
spanned by vectors which are a non-relativistic analogue of the conformal
Killing vectors of flat spacetime. We will take the definition of
these \emph{Schrödinger-Killing} vectors in a general Newton-Cartan
spacetime and proceed to prove the following theorem.

\textcompwordmark{}

\textbf{Theorem 1}

The first-order symmetries of the Schrödinger equation
\begin{equation}
\hat{\Delta}\psi:=i\partial_{t}\psi-\frac{1}{2m}\delta^{jk}\left(-i\partial_{j}+mA_{j}\right)\left(-i\partial_{k}+mA_{k}\right)\psi-mV\psi=0\label{CovSch}
\end{equation}
have the Schrödinger-Killing vectors of the Newton-Cartan spacetime
with Galilean coordinates $(t,x^{i})$ and non-vanishing connection
components
\[
\Gamma_{\,\, tt}^{i}=\delta^{ij}\partial_{j}V\qquad\mbox{and}\qquad\Gamma_{\,\, jt}^{i}=\Gamma_{\,\, tj}^{i}=\delta_{jl}\epsilon^{ilk}\partial_{k}\Omega
\]
as their principal symbols, where $\Omega(x^{j})$ is a function satisfying
$d\Omega=\star^{3}dA$.

\textcompwordmark{}

The operator $\hat{\Delta}$ is the covariant Schrödinger equation
exhibited in \cite{DuvalKunzle}. One can view Theorem 1 as the statement
of a duality between the geometrical properties of a curved spacetime
and the symmetries of an ordinary Schrödinger equation coupled to
potentials.

The non-relativistic analogues of (``conformal'') Killing tensors
will then be discussed in section four, where we will introduce a
Newton-Cartan Hamiltonian formalism in agreement with other approaches
involving Bargmann lifts (see e.g. \cite{DuvalGibbonsHovarthy,GibbonsCarigliaetal}).
The Hamiltonian formalism will then allow us to give natural intrinsic
definitions of Killing tensors and Schrödinger-Killing tensors for
Newton-Cartan geometry, where the latter are a non-relativistic analogue
of conformal Killing tensors. Much like in the Riemannian setting,
the Newton-Cartan Killing tensors correspond to conserved quantities.

We will then proceed to prove the following theorem, a non-relativistic
analogue of Eastwood's identification of the higher symmetries of
the Laplacian as conformal Killing tensors \cite{Eastwood}.

\textcompwordmark{}

\textbf{Theorem 2}

The higher symmetries of the free Schrödinger equation
\[
i\partial_{t}\psi=-\frac{1}{2m}\delta^{ij}\partial_{i}\partial_{j}\psi
\]
are linear differential operators which have the Schrödinger-Killing
tensors of the flat Galilean Newton-Cartan spacetime
\[
h=\delta^{ij}\partial_{i}\partial_{j}\qquad\theta=dt\qquad\Gamma_{\, bc}^{a}=0
\]
as their principal symbols.

\textcompwordmark{}

The higher symmetries of the free Schrödinger equation are well known
\cite{Nikitin}; the novel element here is the correspondence with
a special kind of tensor in Newton-Cartan geometry.

In section 5 we will discuss the links between non-relativistic symmetries
and the Newtonian twistor theory introduced in \cite{DG15}, where
Newton-Cartan geometry in $(3+1)$ dimensions is constructed on the
moduli space of a family of rational curves in a complex manifold
$PT_{\infty}={\cal O}\oplus{\cal O}(2)$, the total space of a rank-two
holomorphic vector bundle over $\mathbb{CP}^{1}$. In particular we
will prove the following theorem, giving a twistorial answer to the
question of what is the non-relativistic analogue of a conformal Killing
vector.

\textcompwordmark{}

\textbf{Theorem 3}

The global holomorphic sections of $T(PT_{\infty})$ are in one-to-one
correspondence with elements of $\mathfrak{cnc}(3)$, a Lie algebra
of vector fields preserving Newton-Cartan geometry with $h=\delta^{ij}\partial_{i}\partial_{j}$
and $\theta=dt$ on $M$. 

\textcompwordmark{}

Both $\mathfrak{cnc}(3)$ and $H^{0}(PT_{\infty},T(PT_{\infty}))$
are infinite-dimensional Lie algebras, and the former was introduced
in \cite{DuvalHovarthy}. The significance of this result comes from
its relativistic counterpart, where the global sections of the twistor
space's tangent bundle are in one-to-one correspondence with the conformal
Killing vectors of the spacetime \cite{Laffar}. We will then proceed
to discuss two subalgebras of $H^{0}(PT_{\infty},T(PT_{\infty}))$,
the expanded Schrödinger algebra and the CGA \cite{Lukierski}.

\section{Newton-Cartan Geometry}

Newton-Cartan spacetimes are the non-relativistic analogues of Lorentzian
manifolds in general relativity: they are the geometrical setting
for non-relativistic physics \cite{Cartan}. Just like in general
relativity we have a four-dimensional manifold playing the role of
the spacetime, and particles travel on geodesics of a torsion-free
connection. There's a metric too, though unlike in general relativity
the connection and the metric are independent quantities. 

\textcompwordmark{}

\textbf{Definition}

A \emph{Newton-Cartan spacetime} (NC) is a quadruplet $(M,\, h,\,\theta,\,\nabla)$
where
\begin{itemize}
\item $M$ is a $(d+1)$-dimensional manifold;
\item $h$ is a symmetric tensor field of valence $\begin{pmatrix}2\\
0
\end{pmatrix}$ with signature $(0++...+)$ (so rank $d$) called the\emph{ metric};
\item $\theta$ is a closed one-form spanning the kernel of $h$ called
the\emph{ clock};
\item and $\nabla$ is a torsion-free connection satisfying $\nabla h=0$
and $\nabla\theta=0$.
\end{itemize}
\textcompwordmark{}

We \foreignlanguage{british}{emphasise} that $\nabla$ must be specified
independently of the metric and clock. Since $\theta$ is closed we
can always locally write $\theta=dt$ for some function $t:M\rightarrow\mathbb{R}$.
This function is then taken as a coordinate on the time axis, a one-dimensional
submanifold over which the NC is fibred. We call the fibres \emph{spatial
slices} and when restricted to such a slice the metric $h$ is a more
familiar signature $(+...++)$ $d$-metric. Throughout this paper
the indices $a,b,c$ will run from $0$ to $d$.

The field equations for NC gravity arise as the Newtonian limit of
the Einstein equations \cite{Kunzle}. They are
\[
R_{ab}=4\pi G\rho\theta_{a}\theta_{b}
\]
where $R_{ab}$ is the Ricci tensor associated to $\nabla$; $G$
is Newton's constant; and $\rho:M\rightarrow\mathbb{R}$ is the mass
density. Alongside the field equations we have the Trautman condition
\cite{DuvalHovarthy}
\begin{equation}
h^{a[b}R_{\,\,(de)a}^{c]}=0\label{TrautmanCondition}
\end{equation}
for $R_{\,\, bcd}^{a}$ the Riemann tensor of $\nabla$. This ensures
that there always exist potentials for the connection components,
which is needed if we are to make contact with Newtonian physics;
accordingly connections which satisfy (\ref{TrautmanCondition}) are
referred to as \emph{Newtonian} connections.

The field equations imply that $h$ is flat on spatial slices, so
we can always introduce \emph{Galilean} coordinates $(t,x^{i})$ such
that
\[
h=\delta^{ij}\frac{\partial}{\partial x^{i}}\otimes\frac{\partial}{\partial x^{j}}\qquad\mbox{and}\qquad\theta=dt
\]
for $i=1,2,...,d$. For notational convenience we can then raise and
lower purely spatial indices with $\delta^{ij}$ and $\delta_{ij}$.

Only connections compatible with $\theta$ and $h$ are allowed; one
can show \cite{Kunzle2} that the most general such connection has
components
\[
\Gamma_{\,\, bc}^{a}=\frac{1}{2}h^{ad}\left(\partial_{b}h_{cd}+\partial_{c}h_{bd}-\partial_{d}h_{bc}\right)+\partial_{(b}\theta_{c)}U^{a}+\theta_{(b}F_{c)d}h^{ad}
\]
where
\begin{itemize}
\item $U^{a}$ is any vector field satisfying $\theta(U)=1$;
\item $F_{ab}$ is any two-form;
\item and $h_{ab}$ is uniquely determined by $h^{ab}h_{bc}=\delta_{c}^{a}-\theta_{c}U^{a}$
and $h_{ab}U^{b}=0$.
\end{itemize}
The possible connections are then parametrised by a choice of $(U,\, F)$.
The Trautman condition (\ref{TrautmanCondition}) is equivalent to
the statement that $F$ is closed, and hence for a Newtonian connection
we can write $F=dA$. Thus we will in future refer to a Newton-Cartan
spacetime as a quintuplet $(M,h,\theta,U,A)$, implicitly considering
a Newtonian connection. Clearly there is a gauge symmetry in $A$,
as we can always shift
\[
A\,\longmapsto\, A+d\chi
\]
for any function $\chi$ on $M$.

There is a further redundancy in this description; there exist \emph{Milne
boosts }which can be thought of as gauge transformations of $(U,\, F)$
which leave $\Gamma_{\,\, bc}^{a}$ unchanged \cite{DuvalHovarthy}.
Usually we will gauge-fix to $U=\partial_{t}$, which can be implemented
for any initial choice of $(U,F)$.

In $d=3$ the most general vacuum Newton-Cartan spacetime satisfying
(\ref{TrautmanCondition}) then has
\[
\Gamma_{\,\, tt}^{i}=\delta^{ij}\partial_{j}V\qquad\mbox{and}\qquad\Gamma_{\,\, jt}^{i}=\Gamma_{\,\, tj}^{i}=\delta_{jl}\epsilon^{ilk}\partial_{k}\Omega
\]
\[
\mbox{where}\qquad\delta^{ij}\partial_{i}\partial_{j}V+2\delta^{ij}\partial_{i}\Omega\partial_{j}\Omega=0\qquad\mbox{and}\qquad\delta^{ij}\partial_{i}\partial_{j}\Omega=0,
\]
with all other connection components vanishing. For reference, the
corresponding two-form $F$ is given by
\[
F=-dV\wedge dt\,+\,\epsilon_{ijk}\delta^{kl}\partial_{l}\Omega\, dx^{i}\wedge dx^{j}.
\]
The geodesic equations suggest that we should interpret the function
$V$ as the Newtonian (gravitational) potential and the function $\Omega$
as a potential for generalised (spatially-varying) Coriolis forces.

\section{First-Order Symmetries and Killing Vectors}

\subsection{Review: the Schrödinger Algebra}

A menagerie of symmetry algebras relevant in Newton-Cartan geometry
is discussed in \cite{DuvalHovarthy}; we will here provide a brief
review of those relevant to this work, following that paper.

\textcompwordmark{}

\textbf{Definition} 

The \emph{expanded Schrödinger algebra} $\widetilde{\mathfrak{sch}}(d)$
(for the flat case) is the Lie algebra of vector fields which conformally
preserve the metric and clock in the sense that
\begin{equation}
\mathcal{L}_{X}h^{ab}=fh^{ab}\label{schmetric}
\end{equation}
\begin{equation}
\mathcal{L}_{X}\theta_{a}=g\theta_{a}\label{schclock}
\end{equation}

and effect projective transformations of the connection such that
\begin{equation}
\mathcal{L}_{X}\Gamma_{\, bc}^{a}=\delta_{(b}^{a}\phi_{c)}\,\,,\label{schconnection}
\end{equation}
with functions $(f,g)$ and a one-form $\phi_{a}$ constrained by
\begin{equation}
\mathcal{L}_{X}\nabla h=0\quad\mbox{and}\quad\mathcal{L}_{X}\nabla\theta=0,\label{schconstraints}
\end{equation}
and where
\begin{equation}
h=\delta^{ij}\partial_{i}\partial_{j}\qquad\theta=dt\qquad\Gamma_{\, bc}^{a}=0\label{GenuinelyFlatNC}
\end{equation}
is the flat Newton-Cartan spacetime.

\textcompwordmark{}

Condition (\ref{schconnection}) ensures that the unparametrised geodesics
are unaltered by the transformation, whilst (\ref{schmetric}) and
(\ref{schclock}) are the non-relativistic analogue of the conformal
Killing equations. As an aside we note that the infinite-dimensional
algebra of vector fields obeying only (\ref{schmetric}-\ref{schclock})
is known as $\mathfrak{cgal}(d)$, the \emph{conformal Galilean algebra}.

We can solve the system (\ref{schmetric}-\ref{schconstraints}) for
the NC (\ref{GenuinelyFlatNC}) and find that $X\in\widetilde{\mathfrak{sch}}(d)$
iff
\begin{equation}
X=\left(\alpha t^{2}+\beta t+\gamma\right)\partial_{t}+\left(\omega_{\, j}^{i}x^{j}+\alpha tx^{i}+\mu x^{i}+\nu^{i}t+\rho^{i}\right)\partial_{i}\label{ExpandedSchX}
\end{equation}
for $(\alpha,\beta,\gamma,\mu,\nu^{i},\rho^{i})\in\mathbb{R}^{4+2d}$
and $\delta_{ik}\omega_{\, j}^{i}=\omega_{kj}\in\mathfrak{so}(d)$.
The dimension of $\widetilde{\mathfrak{sch}}(d)$ is therefore $\frac{1}{2}\left(d^{2}+3d+8\right)$.

\textcompwordmark{}

\textbf{Definition}

The \emph{Schrödinger algebra} $\mathfrak{sch}(d)$ is the Lie subalgebra
of $\widetilde{\mathfrak{sch}}(d)$ defined by the additional condition
$f+g=0$.

\textcompwordmark{}

This amounts to setting $\beta=2\mu$ in (\ref{ExpandedSchX}); we
thus have that $X\in\mathfrak{sch}(d)$ iff
\begin{equation}
X=\left(\alpha t^{2}+2\mu t+\gamma\right)\partial_{t}+\left(\omega_{\, j}^{i}x^{j}+\alpha tx^{i}+\mu x^{i}+\nu^{i}t+\rho^{i}\right)\partial_{i}.\label{SchX}
\end{equation}
Physically, this algebra contains translations $(\gamma,\rho^{i})$,
spatial rotations $(\omega_{\, j}^{i})$, boosts $(\nu^{i})$, a special-conformal
transformation $(\alpha)$, and a dilation $(\mu)$. The dimension
is now $\frac{1}{2}\left(d^{2}+3d+6\right)$.

\textcompwordmark{}

This algebra is named ``Schrödinger'' because of its well-known
link (see e.g. \cite{Son}) to the \emph{free-particle} Schrödinger
equation: a first-order linear differential operator $\mathcal{D}=S^{a}(x)\partial_{a}+s(x)$
commutes with $\Delta=i\partial_{t}+\frac{1}{2m}\delta^{ij}\partial_{i}\partial_{j}$
in the sense that
\[
\Delta\mathcal{D}=\delta\Delta
\]
for some linear differential operator $\delta$ iff $S^{a}\partial_{a}\in\mathfrak{sch}(d)$.

In the remainder of this section we will generalise this statement,
proving Theorem 1.

\subsection{Schrödinger-Killing Vectors on Curved Spacetimes}

The equations (\ref{schmetric}-\ref{schconstraints}) defining the
expanded Schrödinger algebra $\widetilde{\mathfrak{sch}}(d)$ make
sense for a curved Newton-Cartan spacetime as well as a flat one:
we simply use
\begin{equation}
h=\delta^{ij}\partial_{i}\partial_{j}\qquad\theta=dt\qquad\Gamma_{\, tt}^{i}=\delta^{ij}\partial_{j}V\qquad\Gamma_{\,\, jt}^{i}=\Gamma_{\,\, tj}^{i}=\delta_{jl}\epsilon^{ilk}\partial_{k}\Omega\label{NCwithPotentialandCor}
\end{equation}
with all other connection components vanishing instead of (\ref{GenuinelyFlatNC}).
In the following definition we will bypass the \emph{expanded} version
of these vectors and impose the $f+g=0$ constraint from the beginning.

\textcompwordmark{}

\textbf{Definition}

A \emph{Schrödinger-Killing vector} of a curved Newton-Cartan spacetime
(\ref{NCwithPotentialandCor}) is a vector field $X$ obeying (\ref{schmetric}-\ref{schconstraints})
and $f+g=0$.

\textcompwordmark{}

In order to prove Theorem 1 it will be useful to write out in more
detail the equations (\ref{schmetric}-\ref{schconstraints}) on (\ref{NCwithPotentialandCor}).
Thus we collect for reference
\begin{equation}
\partial_{i}X^{t}=0\label{SK1}
\end{equation}
\begin{equation}
\partial^{i}X^{j}+\partial^{j}X^{i}=\partial_{t}X^{t}\delta^{ij}
\end{equation}
\begin{equation}
\partial_{t}\partial_{t}X^{i}+X^{j}\partial_{j}\partial^{i}V+2\partial^{i}V\partial_{t}X^{t}+2\epsilon_{\, jk}^{i}\partial^{k}\Omega\partial_{t}X^{j}-\partial^{j}V\partial_{j}X^{i}=0
\end{equation}
\begin{equation}
\epsilon_{ijk}\partial^{j}\partial_{t}X^{i}+2X^{j}\partial_{j}\partial_{k}\Omega+2\partial_{k}\Omega\partial_{t}X^{t}+\partial_{k}X^{j}\partial_{j}\Omega-\partial^{j}X_{k}\partial_{j}\Omega=0.\label{SK4}
\end{equation}
(Recall that spatial indices are raised and lowered throughout with
Kronecker deltas.)

\textcompwordmark{}

\textbf{Example}

Take the $(3+1)$-dimensional Newton-Cartan spacetime with the linear
Newtonian potential $V=z$, adopting $x^{i}=(x,y,z)$. The Riemann
tensor vanishes, so we expect the symmetry group to be of maximal
dimension. Solving (\ref{SK1}-\ref{SK4}) yields
\begin{multline*}
X=\left(\alpha t^{2}+2\mu t+\gamma\right)\partial_{t}+\left(\omega_{\, j}^{i}x^{j}+\alpha tx^{i}+\mu x^{i}+\nu^{i}t+\rho^{i}\right)\partial_{i}\\
+\frac{1}{2}\omega^{xz}t^{2}\partial_{x}+\frac{1}{2}\omega^{yz}t^{2}\partial_{y}-\left(\frac{2}{3}\alpha t^{3}+2\mu t^{2}\right)\partial_{z}.
\end{multline*}
We thus indeed find a twelve-dimensional algebra, though the vectors
come with some additional terms which result from the strange choice
of coordinates.

\textcompwordmark{}

\textbf{Example}

The Schrödinger-Killing vectors of the $(3+1)$-dimensional Newton-Cartan
spacetime with $V=(x^{2}+y^{2}+z^{2})^{-\frac{1}{2}}$ and $\Omega=0$
are
\[
X=\gamma\partial_{t}+\omega_{\, j}^{i}x^{j}\partial_{i}
\]
for $\gamma$ a constant and $\omega_{jk}\in\mathfrak{so}(3)$. The
presence of a point mass at the origin has reduced the symmetry algebra
to just time translations and spatial rotations.

\subsection{Symmetries of the Covariant Schrödinger Operator}

In this subsection we will consider the first-order symmetries of
the operator
\[
\hat{\Delta}=i\partial_{t}-\frac{1}{2m}\delta^{jk}\left(-i\partial_{j}+mA_{j}\right)\left(-i\partial_{k}+mA_{k}\right)-mV,
\]
where $V$ and $A_{i}$ depend on space only. That is to say, we will
seek first-order linear differential operators
\[
\mathcal{D}=S^{a}(x^{b})\partial_{a}+s(x^{b})
\]
which obey 
\begin{equation}
\hat{\Delta}\mathcal{D}=\delta\hat{\Delta}\label{commute}
\end{equation}
for some (otherwise irrelevant) linear differential operator $\delta$.

\textcompwordmark{}

\textbf{Proof of Theorem 1}

If we calculate the left-hand-side of (\ref{commute}) then we get
$\mathcal{D}\hat{\Delta}$, which is already in the right form, and
some additional operator terms. These additional terms arrange themselves
into $\hat{\Delta}$ iff
\begin{equation}
\partial_{i}S^{t}=0\label{SYM1}
\end{equation}
\begin{equation}
\partial^{i}S^{j}+\partial^{j}S^{i}=\delta^{ij}\partial_{t}S^{t}\label{SYM2}
\end{equation}
\begin{equation}
-im\partial_{t}S^{i}-imA^{j}\partial_{j}S^{i}+imS^{j}\partial_{j}A^{i}+imA^{i}\partial_{t}S^{t}=\partial^{i}s\label{SYM3}
\end{equation}
\begin{equation}
\frac{i}{2m}\partial_{i}\partial^{i}s-A^{i}\partial_{i}s-\left(iS^{j}\partial_{j}+i\partial_{t}S^{t}\right)\left(\frac{i}{2}\partial_{i}A^{i}-\frac{m}{2}A^{i}A_{i}-mV\right)=\partial_{t}s.\label{SYM4}
\end{equation}
In order to prove Theorem 1 we must find the conditions on $S^{a}$
such that one can always find $s$ solving these equations. To that
end we use (\ref{SYM3}) to rewrite $\frac{i}{2m}\partial_{i}\partial^{i}s-A^{i}\partial_{i}s$
in (\ref{SYM4}) in terms of $S^{a}$ only. Then (\ref{SYM3}-\ref{SYM4})
have the form
\[
\Sigma=ds
\]
for $\Sigma_{a}=\Sigma_{a}(S^{b},A^{i},V)$. By the Poincaré lemma
the conditions we are looking for are
\begin{equation}
d\Sigma=0.\label{dSigmaequalsZero}
\end{equation}
Explicit calculation reveals that (\ref{SYM1},\ref{SYM2},\ref{dSigmaequalsZero})
are then exactly the equations (\ref{SK1}-\ref{SK4}) defining Schrödinger-Killing
vectors with $d\Omega=\star^{3}dA$, completing the proof of Theorem
1.\hfill{}$\square$

\textcompwordmark{}

Note that the gauge symmetry
\[
A_{i}\,\longmapsto\, A_{i}+\partial_{i}\chi
\]
has not here been fixed. The Schrödinger-Killing vectors of the curved
NC spacetime are the symmetries of the whole gauge equivalence class
of operators $\hat{\Delta}$.

\section{Higher Symmetries and Killing Tensors}

\subsection{Non-Relativistic Killing Tensors and Conserved Quantities}

In this subsection we will define the non-relativistic analogues of
Killing tensors by exhibiting Newton-Cartan geodesics as the projection
of the integral curves of a Hamiltonian vector field on the cotangent
bundle. This Hamiltonian formalism is an intrinsic Newton-Cartan analogue
of the Eisenhart-Duval lift discussed in, say, \cite{GibbonsCarigliaetal}.

\textcompwordmark{}

\textbf{Lemma 1}

Geodesics of the Newton-Cartan spacetime $(M,h,\theta,F)$ with connection
components
\begin{equation}
\qquad\Gamma_{\, bc}^{a}=\frac{1}{2}h^{ad}\left(\partial_{b}h_{cd}+\partial_{c}h_{bd}-\partial_{d}h_{bd}\right)+\partial_{(b}\theta_{c)}U^{a}+\theta_{(b}F_{c)d}h^{ad}\label{GenericNewtonNC}
\end{equation}
and with $F=dA$ are the projection from $T^{*}M$ to $M$ of the
integral curves of the geodesic spray
\begin{gather*}
\mathcal{G}=\left(\frac{1}{2}\partial_{a}h^{cd}\Pi_{c}\Pi_{d}+h^{cd}\Pi_{c}\partial_{a}A_{d}-\partial_{a}U^{b}\Pi_{b}-U^{b}\partial_{a}A_{b}\right)\frac{\partial}{\partial p_{a}}+\left(U^{a}-h^{ab}\Pi_{b}\right)\frac{\partial}{\partial x^{a}}
\end{gather*}
(where $\Pi_{a}:=p_{a}+A_{a}$ and $(x^{a},p_{b})\in T^{*}M$), which
is the Hamiltonian vector field associated to
\[
\mathcal{H}=\frac{1}{2}h^{ab}\Pi_{a}\Pi_{b}-U^{a}\Pi_{a}.
\]

The proof of this lemma is straightforward (but tedious); we omit
it for brevity.

\textcompwordmark{}

This Hamiltonian (and therefore also the following definitions) are
Milne-boost invariant.

\textcompwordmark{}

\textbf{Definition}

A rank-$n$ \emph{Killing tensor }of a Newton-Cartan spacetime $(M,h,\theta,U,F)$
is a symmetric contravariant tensor field $X^{a_{1}...a_{n}}$ such
that functions $\chi_{m}^{a_{1}...a_{m}}$ on $M$ can be found obeying
\begin{equation}
\left\{ X^{a_{1}...a_{n}}p_{a_{1}}...p_{a_{n}}+\sum_{m=0}^{n-1}\chi_{m}^{a_{1}...a_{m}}p_{a_{1}}...p_{a_{m}}\,,\,\mathcal{H}\right\} =0\,\,,\label{PoissonwithH}
\end{equation}
where $\left\{ \,,\,\right\} $ is the canonical Poisson structure
on $T^{*}M$. The quantity 
\[
X^{a_{1}...a_{n}}p_{a_{1}}...p_{a_{n}}+\sum_{m=0}^{n-1}\chi_{m}^{a_{1}...a_{m}}p_{a_{1}}...p_{a_{m}}
\]
is constant along geodesics.

Here we have provided an intrinsic Newton-Cartan definition of the
usual concept of a hidden symmetry, entirely in line with the familiar
concept from classical dynamics.

\textcompwordmark{}

Taking $n=1$ in (\ref{PoissonwithH}) we arrive at the conditions
\begin{equation}
\mathcal{L}_{X}h=0\label{NCKV1}
\end{equation}
\begin{equation}
\mathcal{L}_{X}U-h\left(\mathcal{L}_{X}A\,,\,\,\right)=-h\left(d\chi_{0},\,\,\right)\label{NCKV2-1}
\end{equation}
\begin{equation}
\left(\mathcal{L}_{X}A\right)(U)=d\chi_{0}(U).\label{NCKV2}
\end{equation}
Solving (\ref{NCKV1}-\ref{NCKV2}) on a given Newton-Cartan spacetime
will give us the Killing vectors of that spacetime.

\textcompwordmark{}

\textbf{Example}

$X=X^{a}\partial_{a}$ solves (\ref{NCKV1}-\ref{NCKV2}) with
\begin{equation}
h=\delta^{ij}\partial_{i}\partial_{j}\quad\theta=dt\quad U=\partial_{t}\quad A=0\label{FGNC}
\end{equation}
and is thus a non-relativistic Killing vector of the flat Newton-Cartan
spacetime iff
\[
X=\gamma\partial_{t}+\left(\omega_{\, j}^{i}x^{j}+\nu^{i}t+\rho^{i}\right)\partial_{i}
\]
for any ten constants $(\gamma,\nu^{i},\rho^{i},\,\omega_{ij}\in\mathfrak{so}(3)$).
Such vectors generate the Galilean group.

\textcompwordmark{}

\textbf{Example}

The $(3+1)$-dimensional Newton-Cartan spacetime
\[
h=\delta^{ij}\partial_{i}\partial_{j}\quad\theta=dt\quad U=\partial_{t}\quad A=-(\delta_{lk}x^{l}x^{k})^{-\frac{1}{2}}dt
\]
\foreignlanguage{british}{modelling} the Kepler problem (where $b$
is a constant) admits the following three rank-two non-relativistic
Killing tensors
\[
X^{ij}=\lambda^{l}x^{k}\delta_{lk}\delta^{ij}-\lambda^{(i}x^{j)}\qquad X^{it}=X^{tt}=0
\]
(for $\lambda^{i}\in\mathbb{R}^{3}$). The lower order terms are
\[
\chi_{1}^{a}=0\qquad\mbox{and}\qquad\chi_{0}=\frac{\lambda^{i}\delta_{ij}x^{j}}{(\delta_{lk}x^{l}x^{k})^{\frac{1}{2}}}\,,
\]
and the three associated conserved quantities together form the famous
Laplace\textendash{}Runge\textendash{}Lenz vector (see e.g. \cite{Cariglia}).

\textcompwordmark{}

\subsection{Schrödinger-Killing Tensors}

In generalising the Schrödinger algebra $\mathfrak{sch}(d)$ to the
case of Schrödinger-Killing tensors we will again make use of the
Hamiltonian formalism introduced above. The following definition is,
in the Hamiltonian formalism, a natural way to define a notion of
a conformal Killing tensor.

\textcompwordmark{}

\textbf{Definition}

A \emph{Schrödinger-Killing tensor} of a Newton-Cartan spacetime $(M,h,\theta,U,F)$
is a symmetric contravariant tensor field $X^{a_{1}...a_{n}}$ for
which functions $\chi_{m}^{a_{1}...a_{m}}$ on $M$ can be found obeying

\begin{equation}
\left\{ X^{a_{1}...a_{n}}p_{a_{1}}...p_{a_{n}}+\sum_{m=0}^{n-1}\chi_{m}^{a_{1}...a_{m}}p_{a_{1}}...p_{a_{m}}\,,\,\mathcal{H}\right\} =\sum_{m=0}^{n-1}\left(f_{m}^{a_{1}...a_{m}}p_{a_{1}...}p_{a_{m}}\right)\mathcal{H}\,\,,\label{SKtDef}
\end{equation}
where $f_{m}^{a_{1}...a_{m}}$ are symmetric tensor fields determined
in terms of $\left(X^{a_{1}...a_{n}},\chi_{m}^{a_{1}...a_{m}}\right)$.

\textcompwordmark{}

A Killing tensor as defined above is a special case of a Schrödinger-Killing
tensor.

If $n=1$ we have
\begin{equation}
\mathcal{L}_{X}h=f_{0}h\label{NCKV1-1}
\end{equation}
\begin{equation}
\mathcal{L}_{X}U-h\left(\mathcal{L}_{X}A\,,\,\,\right)=f_{0}U-h\left(d\chi_{0},\,\,\right)\label{NCKV2-1-1}
\end{equation}
\begin{equation}
\left(\mathcal{L}_{X}A\right)(U)=d\chi_{0}(U).\label{NCKV2-2}
\end{equation}
Using the flat Newton-Cartan spacetime (\ref{FGNC}) reduces this
definition to that of $\mathfrak{sch}(d)$ above.

\textcompwordmark{}

In order to prove Theorem 2 we will display in more detail the conditions
describing the Schrödinger-Killing tensors of the flat NC. The defining
condition (\ref{SKtDef}) becomes the coupled family of equations
\begin{equation}
-\partial^{i}X^{a_{1}...a_{n}}p_{i}p_{a_{1}}...p_{a_{n}}=\frac{1}{2}\delta^{ij}f_{n-1}^{a_{1}...a_{n-1}}p_{i}p_{j}p_{a_{1}}...p_{a_{n-2}}\label{Schouten with p}
\end{equation}
\begin{equation}
\partial_{t}X^{a_{1}...a_{n}}p_{a_{1}}...p_{a_{n}}-\partial^{i}\chi_{n-1}^{a_{1}...a_{n-1}}p_{i}p_{a_{1}}...p_{a_{n-1}}=-f_{n-1}^{a_{1}...a_{n-1}}p_{t}p_{a_{1}}...p_{a_{n-1}}+\frac{1}{2}\delta^{ij}f_{n-2}^{a_{1}...a_{n-2}}p_{i}p_{j}p_{a_{1}}...p_{a_{n-2}}
\end{equation}
\begin{equation}
\partial_{t}\chi_{n-1}^{a_{1}...a_{n-1}}p_{a_{1}}...p_{a_{n-1}}-\partial^{i}\chi_{n-2}^{a_{1}...a_{n-2}}p_{i}p_{a_{1}}...p_{a_{n-2}}=-f_{n-2}^{a_{1}...a_{n-2}}p_{t}p_{a_{1}}...p_{a_{n-2}}+\frac{1}{2}\delta^{ij}f_{n-3}^{a_{1}...a_{n-3}}p_{i}p_{j}p_{a_{1}}...p_{a_{n-3}}
\end{equation}
\[
\vdots
\]
\begin{equation}
\partial_{t}\chi_{2}^{a_{1}a_{2}}p_{a_{1}}p_{a_{2}}-\partial^{i}\chi_{1}^{a_{1}}p_{i}p_{a_{1}}=-f_{1}^{a_{1}}p_{t}p_{a_{1}}+\frac{1}{2}\delta^{ij}f_{0}p_{i}p_{j}
\end{equation}
\begin{equation}
\partial_{t}\chi_{1}^{a_{1}}p_{a_{1}}-\partial^{i}\chi_{0}p_{i}=-f_{0}p_{t}
\end{equation}
\begin{equation}
\partial_{t}\chi_{0}=0\,\,.\label{FinalwithpSchouten}
\end{equation}
We can rewrite these concisely using the Schouten brackets of $X$
with $h$ and $U$, denoted $\mathcal{L}_{X}h$ and $\mathcal{L}_{X}U$.
They become
\begin{equation}
\mathcal{L}_{X}h=f_{n-1}h\label{SchoutenSKt1}
\end{equation}
\begin{equation}
\mathcal{L}_{\chi_{n-1}}h-2\mathcal{L}_{X}U=f_{n-2}h-2f_{n-1}U\label{NextSKESchouten}
\end{equation}
\begin{equation}
\mathcal{L}_{\chi_{n-2}}h-2\mathcal{L}_{\chi_{n-1}}U=f_{n-3}h-2f_{n-2}U\,\,\label{SchoutenSKt2}
\end{equation}
\[
\vdots\,\,\mbox{et.c.}
\]
with this pattern continuing on the understanding that for negative
$m$ we have $f_{m}=\chi_{m}=0$, and where all indices on the right-hand-side
products are symmetrised.

\subsection{Higher Symmetry Operators}

The higher symmetries of the Laplacian and of various Schrödinger
operators have been calculated and are to be found in the literature
\cite{Bekaert,Eastwood,Nikitin}. In this subsection we will define
such symmetries, following those papers, and then proceed to prove
Theorem 2, identifying the higher symmetries of the free Schrödinger
operator with the Schrödinger-Killing tensors of the flat Newton-Cartan
spacetime.

\subsection*{The Laplacian}

In \cite{Eastwood} Eastwood finds the \emph{higher symmetries} of
the Laplacian. These are linear differential operators
\[
\mathcal{D}=V_{n}^{\mu_{1}...\mu_{n}}\frac{\partial^{n}}{\partial x^{\mu_{1}}\partial x^{\mu_{2}}...\partial x^{\mu_{n}}}+V_{n-1}^{\mu_{1}...\mu_{n-1}}\frac{\partial^{n-1}}{\partial x^{\mu_{1}}\partial x^{\mu_{2}}...\partial x^{\mu_{n-1}}}+...+V_{1}^{\mu_{1}}\frac{\partial}{\partial x^{\mu_{1}}}+V_{0}
\]
which commute with the Laplacian $\Delta_{L}$ in the sense that
\[
\Delta_{L}\mathcal{D}=\delta\Delta_{L}
\]
for some linear differential operator $\delta$ (determined by $\mathcal{D}$).
The functions $V_{p}^{\mu_{1}...\mu_{p}}$ (for $0\leq p\leq n$)
are the components of totally symmetric rank-$p$ tensor fields on
flat spacetime, and the tensor of highest rank is called the \emph{symbol}
of the symmetry operator. Eastwood finds that if $\mathcal{D}$ is
a symmetry of the Laplacian then its symbol is a conformal Killing
tensor on flat spacetime, i.e.
\begin{equation}
\partial^{(\mu_{0}}V_{n}^{\mu_{1}...\mu_{n})}=g^{(\mu_{0}\mu_{1}}k^{\mu_{2}...\mu_{n})}\label{CKe}
\end{equation}
for some rank-$(n-1)$ tensor field $k$ (which itself is determined
from (\ref{CKe})) and inverse (flat) metric $g^{\mu\nu}$. Furthermore,
when (\ref{CKe}) is satisfied for some symbol $V_{n}^{\mu_{1}...\mu_{n}}$
one can uniquely solve for lower order operators $(V_{n-1}^{\mu_{1}...\mu_{n-1}},V_{n-2}^{\mu_{1}...\mu_{n-2}},...,V_{0})$
determined in terms of the symbol such that ${\cal D}$ is a symmetry
of the Laplacian.

\subsection*{The Free Schrödinger Operator}

The analogous higher symmetries of the free-particle Schrödinger operator
\[
\Delta=i\partial_{t}+\frac{1}{2m}\delta^{ij}\partial_{i}\partial_{j}
\]
can be found in the literature \cite{Nikitin}. Here we will summarise
and make use of the approach of \cite{Bekaert}, where the symmetries
of $\Delta$ in $d+1$ dimensions arise as the light-cone reduction
of conformal Killing tensors in $d+2$ dimensions. 

\textcompwordmark{}

Consider the wave equation in $d+2$ dimensions, written in light-cone
coordinates $(x^{i},x^{+},x^{-})$:
\[
\Delta_{L}\phi=\left(\delta^{ij}\partial_{i}\partial_{j}-2\partial_{+}\partial_{-}\right)\phi=0.
\]
Restricting to fields of the form
\begin{equation}
\phi(x^{i},x^{+},x^{-})=\psi(x^{+},x^{i})\exp\left\{ -imx^{-}\right\} \label{ansatz}
\end{equation}
reduces the wave equation to
\[
\left(i\partial_{+}+\frac{1}{2m}\delta^{ij}\partial_{i}\partial_{j}\right)\psi(x^{+},x^{i})=0\,\,,
\]
which is just $\Delta\psi=0$ if we identify $x^{+}$ with time.

Let $\mathcal{D}$ be a symmetry of the Laplacian, allowing us to
write
\[
\Delta_{L}\mathcal{D}\phi=\delta\Delta_{L}\phi.
\]
Restricting to the ansatz (\ref{ansatz}) reduces this to
\begin{equation}
\Delta_{L}{\cal D}\left(e^{-imx^{-}}\psi\right)=\delta e^{-imx^{-}}\Delta\phi.\label{intermediatansatz}
\end{equation}
Applying ${\cal D}$ to $e^{-imx^{-}}\psi$ results in a new symmetry
operator $\tilde{D}$:
\[
\mathcal{D}\left(e^{-imx^{-}}\psi\right)=e^{-imx^{-}}\tilde{{\cal D}}\psi.
\]
The left-hand-side of (\ref{intermediatansatz}) rearranges into $\Delta$
iff $\partial_{-}\tilde{D}=0$, giving us
\[
\Delta\tilde{{\cal D}}\psi=\tilde{\delta}\Delta\psi\qquad\mbox{for}\,\,\,\partial_{-}\tilde{D}=0.
\]
We can reverse these steps, giving us the statement that the higher
symmetries of $\Delta$ are the operators $\tilde{{\cal D}}$, arising
from conformal Killing tensors.

\textcompwordmark{}

\textbf{Proof of Theorem 2}

To prove Theorem 2 we will consider the conformal Killing equation
in $d+2$ dimensions with coordinates $x^{\mu}=(x^{i},x^{+},x^{-})$.
We will calculate the resulting conditions on ${\cal D}$ and compare
them to the equations (\ref{Schouten with p}-\ref{FinalwithpSchouten})
characterising Schrödinger-Killing tensors. 

For a tensor of rank $n$ the conformal Killing equation is 
\[
\partial^{(\mu_{0}}S^{\mu_{1}...\mu_{n})}=g^{(\mu_{0}\mu_{1}}k^{\mu_{2}...\mu_{n})}
\]
where the only non-vanishing components of the metric are
\[
g_{ij}=\delta_{ij}\qquad g_{+-}=g_{-+}=-1\,.
\]
Recall that we are only interested in solutions not-depending on $x^{-}$.

Consider first the case $(\mu_{0}...\mu_{n})=(a_{0}...a_{n})$, i.e.
no $(-)$ indices are included. We can then identify
\[
g^{ab}=h^{ab}\qquad\partial^{a}=h^{ab}\partial_{b}\,\,,
\]
giving us
\[
h^{b(a_{0}}\partial_{b}S^{a_{1}...a_{n})}=h^{(a_{0}a_{1}}k^{a_{2}...a_{n})}.
\]
Writing $X^{a_{1}...a_{n}}=S^{a_{1}...a_{n}}$ and $k^{a_{1}...a_{n-1}}=-\frac{1}{2}f_{n-1}^{a_{1}...a_{n-1}}$,
we then have
\[
\mathcal{L}_{X}h=f_{n-1}h\,\,,
\]
the first of the equations (\ref{Schouten with p}-\ref{FinalwithpSchouten})
characterising the Schrödinger-Killing tensor $X^{a_{1}...a_{n}}$.
Similarly we can start to set one index to $(-)$ and the rest to
$(a_{1}...a_{n})$, giving
\[
\partial^{-}S^{a_{1}...a_{n}}+n\partial^{(a_{1}}S^{a_{2}...a_{n})-}=2g^{-(a_{1}}k^{a_{2}...a_{n})}+(n-1)g^{(a_{1}a_{2}}k^{a_{3}...a_{n})-}
\]
\[
\Rightarrow\,\,\partial_{+}S^{a_{1}...a_{n}}-nh^{b(a_{1}}\partial_{b}S^{a_{2}...a_{n})-}=2\delta_{+}^{(a_{1}}k^{a_{2}...a_{n})}-(n-1)h^{(a_{1}a_{2}}k^{a_{3}...a_{n})-}.
\]
Again, this equation is the same as (\ref{NextSKESchouten}) with
the identifications
\begin{equation}
\chi_{n-1}^{a_{1}...a_{n-1}}=nS^{-a_{1}...a_{n-1}}\qquad f_{n-2}^{a_{1}...a_{n-2}}=-2(n-1)k^{-a_{1}...a_{n-2}}.\label{IdentifyfkSchi}
\end{equation}
In fact, all of the equations (\ref{Schouten with p}-\ref{FinalwithpSchouten})
can be reproduced in this manner from the conformal Killing equation,
one for each number of indices set to $(-)$, and with similar identifications
to (\ref{IdentifyfkSchi}). The equation with $q$ indices set to
$(-)$ and the rest set to $(a_{1}...a_{n+1-q})$ is
\begin{multline*}
q\partial^{-}S^{-...-a_{1}...a_{n+1-q}}+(n+1-q)\partial^{(a_{1}}S^{a_{2}...a_{n+1-q})-...-}\\
=2\frac{q}{n}(n+1-q)g^{-(a_{1}}k^{a_{2}...a_{n+1-q})-...-}+\frac{1}{n}(n+1-q)(n-q)g^{(a_{1}a_{2}}k^{a_{3}...a_{n+1-q})-...-}
\end{multline*}
\begin{multline*}
\Rightarrow\quad q\partial_{+}S^{-...-a_{1}...a_{n+1-q}}-(n+1-q)h^{b(a_{1}}\partial_{b}S^{a_{2}...a_{n+1-q})-...-}\\
=2\frac{q}{n}(n+1-q)\delta_{+}^{(a_{1}}k^{a_{2}...a_{n+1-q})-...-}-\frac{1}{n}(n+1-q)(n-q)h^{(a_{1}a_{2}}k^{a_{3}...a_{n+1-q})-...-}.
\end{multline*}
We can then identify
\[
\chi_{n-q}^{a_{1}...a_{n-q}}=\begin{pmatrix}n\\
q
\end{pmatrix}S^{a_{1}...a_{n-q}-...-}\quad\quad\mbox{for }q\geq1
\]
\[
\mbox{and}\quad f_{n-q}^{a_{1}...a_{n-q}}=-2\begin{pmatrix}n-1\\
q-1
\end{pmatrix}k^{a_{1}...a_{n-q}-...-}\quad\quad\mbox{for }q\geq1.
\]
These equations are now exactly those above characterising Schrödinger-Killing
tensors, completing the proof of Theorem 2. \hfill{}$\square$

\textcompwordmark{}

With this achieved, a natural question to ask is whether this result
extends to curved Newton-Cartan spacetimes and the covariant Schrödinger
equation. The situation here remains unclear, just as it does in the
case of the curved Riemannian manifold and the Laplacian, and we defer
this question to future investigations.

\section{Conformal Symmetries on Newtonian Twistor Space}

\subsection{Review: Newtonian Twistor Theory}

Newton-Cartan spacetimes admit a twistorial construction as the moduli
space of rational curves in the twistor space. The details of this
construction are to be found in \cite{DG15}; here we will provide
a short review in order to discuss the role of conformal Newton-Cartan
symmetries in the associated twistor theory.

Let $(\lambda,\hat{\lambda})$ be inhomogeneous coordinates on two
patches defined by stereographic projection from the north and south
poles of the Riemann sphere $\mathbb{CP}^{1}$, and let
\[
\begin{array}{c}
{\cal O}(n)\\
\downarrow\\
\mathbb{CP}^{1}
\end{array}
\]
be a rank-one holomorphic vector bundle over $\mathbb{CP}^{1}$ with
patching 
\[
\hat{s}=\lambda^{-n}s\,\,,
\]
where $(s,\hat{s})$ are coordinates on the fibres over the two patches.

The twistor space is the total space of the bundle $PT_{\infty}={\cal O}\oplus{\cal O}(2)$,
fibred over $\mathbb{CP}^{1}$. Let $PT_{\infty}$ be covered by two
patches $U$ and $\hat{U}$ with coordinates $Z^{\alpha}=(T,Q,\lambda)$
and $\hat{Z}^{\alpha}=(\hat{T},\hat{Q},\hat{\lambda})$, where $(T,Q)$
are coordinates on the fibres and $\lambda$ is a coordinate on the
base $\mathbb{CP}^{1}$. The patching is then
\[
\hat{T}=T\qquad\hat{Q}=\lambda^{-2}Q\qquad\hat{\lambda}=\lambda^{-1}
\]
We identify the $(3+1)$-dimensional Newton-Cartan spacetime $M$
as the moduli space of global sections of $PT_{\infty}\rightarrow\mathbb{CP}^{1}$,
by means of the double-fibration of the projective spin bundle $P\mathbb{S}^{\prime}$.
\[
\begin{array}{ccccc}
\\
 &  & P\mathbb{S^{\prime}}\\
 & \overset{\nu}{\swarrow} &  & \overset{\mu}{\searrow}\\
M &  &  &  & PT_{\infty}\\
\\
\end{array}
\]

The maps realising the fibrations are
\[
\nu(x^{a},\lambda)=x^{a}
\]
and
\[
\mu:\,(x^{a},\lambda)\,\longmapsto\,\begin{pmatrix}T\\
Q\\
\lambda
\end{pmatrix}=\begin{pmatrix}t\\
\lambda^{2}(x-iy)-2\lambda z-(x+iy)\\
\lambda
\end{pmatrix}.
\]

Global holomorphic data on $PT_{\infty}$ can be mapped to $M$ giving
rise to the Galilean structure. The mapping procedure (for the case
of vector fields) is discussed concretely in section 5.2.
\begin{itemize}
\item $H^{0}(PT_{\infty},T^{*}(PT_{\infty}))$ contains one-forms $k(T)dT$
for holomorphic $k(T)$, which correspond to one-forms $k(t)dt$ on
$M$. This gives us the conformal clock $[\theta]$.
\item The zero of $H^{0}(PT_{\infty},T(PT_{\infty})^{\odot2})$ corresponds
to the conformal structure $[h]$ containing $\delta^{ij}\partial_{i}\partial_{j}$
on $M$.
\end{itemize}
Both of these conformal factors can be fixed using data on the non-projective
twistor space to give $h=\delta^{ij}\partial_{i}\partial_{j}$ and
$\theta=dt$: see \cite{DG15} for details. The construction of the
affine connection on $M$ need not concern us here, and we again refer
the reader to \cite{DG15}.

\subsection{Holomorphic Vector Fields on $PT_{\infty}$ and $\mathfrak{cnc}(3)$}

In the nonlinear graviton construction the conformal symmetries of
the spacetime are in one-to-one correspondence with holomorphic vector
fields on twistor space, that is to say the conformal symmetries on
$M$ arise as global sections of $T(PT)$ \cite{Laffar}.

It is thus natural to ask what the global sections of $T(PT_{\infty})$
correspond to on the Newton-Cartan spacetime. The answer is the following
theorem.

\textcompwordmark{}

\textbf{Theorem 3}

The global sections of $T(PT_{\infty})$ are in one-to-one correspondence
with conformal Newton-Cartan vectors $\mathfrak{cnc}(3)$ of the Galilean
structure $h=\delta^{ij}\partial_{i}\partial_{j}$ and $\theta=dt$
on $M$.

\textcompwordmark{}

The algebra $\mathfrak{cnc}(3)$ is discussed in \cite{DuvalHovarthy},
where it is defined to be the algebra of vector fields on a Newton-Cartan
spacetime which generate conformal transformations of $(h,\theta)$
and null-projective transformations of $\nabla$. Both $\mathfrak{cnc}(3)$
and $H^{0}(PT_{\infty},T(PT_{\infty}))$ are infinite-dimensional
Lie algebras.

We will prove Theorem 3 by directly calculating the holomorphic vector
fields $\beta^{\alpha}$ on $PT_{\infty}$. The patching
\[
\hat{\beta}^{\rho}=\frac{\partial\hat{Z}^{\rho}}{\partial Z^{\sigma}}\beta^{\sigma}
\]
can be expanded to give
\begin{gather*}
\hat{\beta}^{T}=\beta^{T}\\
\hat{\beta}^{Q}=\lambda^{-2}\beta^{Q}-2\lambda^{-3}Q\beta^{\lambda}\\
\hat{\beta}^{\lambda}=-\lambda^{-2}\beta^{\lambda}.
\end{gather*}
By considering an ansatz in which $\beta^{\alpha}$ are arbitrary
polynomials in $(Q,\lambda)$ whose coefficients are arbitrary holomorphic
functions of the trivial coordinate $T$ we find that $\beta\in H^{0}\left(PT_{\infty},T(PT_{\infty})\right)$
iff 
\begin{multline}
\beta=h(T)\frac{\partial}{\partial T}+\left(a(T)+b(T)Q+c(T)\lambda+d(T)\lambda Q+e(T)\lambda^{2}\right)\frac{\partial}{\partial Q}\\
+\left(f(T)+g(T)\lambda+\frac{1}{2}\lambda^{2}d(T)\right)\frac{\partial}{\partial\lambda}\label{H0TPTinfty}
\end{multline}
for $(a,b,c,...,h)$ any eight holomorphic functions of $T$. These
sections form an infinite-dimensional Lie algebra (under the usual
commutator).

Pushing this algebra to $M$ is a two-stage procedure. First we consider
an arbitrary vector $\Lambda\in T(P\mathbb{S}^{\prime})$ and its
push-forward to $PT_{\infty}$:
\[
(\mu_{\star}\Lambda)^{\alpha}=\frac{\partial\left(Z^{\alpha}|\right)}{\partial x^{\Sigma}}\Lambda^{\Sigma},
\]
where $x^{\Sigma}=(x^{a},\lambda)$ are coordinates on $\mu^{-1}(U)\subset P\mathbb{S}^{\prime}$.
Thus setting $\beta^{\alpha}=(\mu_{\star}\Lambda)^{\alpha}$ we have
\[
\beta^{T}=\Lambda^{t}\qquad\beta^{Q}-\frac{\partial(Q|)}{\partial\lambda}\beta^{\lambda}=\Lambda^{i}\frac{\partial(Q|)}{\partial x^{i}}\qquad\beta^{\lambda}=\Lambda^{\lambda},
\]
and we can uniquely determine a vector $\Lambda$ such that $\Lambda^{a}$
does not depend on $\lambda$ (necessary for the next step). The second
half of the procedure is to simply push-down $\Lambda$ to $X=\nu_{\star}\Lambda$
on $M$, giving
\[
X=\Lambda^{a}(x^{i},t)\frac{\partial}{\partial x^{a}}.
\]
Doing this for the general global vector (\ref{H0TPTinfty}) yields
\[
X=h(t)\frac{\partial}{\partial t}+\left(\omega_{\, j}^{i}(t)x^{j}+\chi(t)x^{i}+\eta^{i}(t)\right)\frac{\partial}{\partial x^{i}}
\]
where 
\begin{gather*}
\chi(t)=b(t)-g(t)\\
\omega_{\, y}^{x}(t)=ig(t)\\
\omega_{\, x}^{z}(t)=f(t)+\frac{1}{2}d(t)\\
\omega_{\, y}^{z}(t)=i\left(\frac{1}{2}d(t)-f(t)\right)\\
\eta^{i}(t)\frac{\partial(Q|)}{\partial x^{i}}=a(t)+c(t)\lambda+e(t)\lambda^{2},
\end{gather*}
revealing $X$ to be an arbitrary element of $\mathfrak{cnc}(3)$.
The procedure is reversible with no trouble: the $\partial_{\lambda}$
component of the lift of $X$ to $P\mathbb{S}^{\prime}$ is calculated
by requiring that the resulting vector field should descend to a vector
field on $PT_{\infty}$. This completes the proof of Theorem 3.\hfill{}$\square$

Note that the factors of $i$ above do not prevent $X$ from being
real; it is possible to choose the real and imaginary parts of $(a,b,c,...,h)$
such that $X$ is any element of the real $\mathfrak{cnc}(3)$.

\subsection{The Expanded Schrödinger Algebra on $PT_{\infty}$}

The expanded Schrödinger algebra $\widetilde{\mathfrak{sch}}(3)$
in a finite-dimensional subalgebra of $\mathfrak{cnc}(3)$, and so
it is to natural to ask what characterises the corresponding holomorphic
vector fields on $PT_{\infty}$. On $M$ we pick out the subalgebra
(as described in section three) by a requirement that the vector must
generate projective transformations of a connection. In this section
we will describe an analogous procedure which takes place in twistor
space.

In order to have a notion of a projective holomorphic vector field
on $PT_{\infty}$ we must first establish some kind of affine connection
on the twistor space. Using the standard law for transforming connection
components we can write down the patching for a new bundle $\mathcal{G}\rightarrow PT_{\infty}$,
whose sections are (the components of) torsion-free affine connections
on twistor space. The patching is
\begin{equation}
\hat{\Gamma}_{\,\beta\gamma}^{\alpha}=\frac{\partial\hat{Z}^{\alpha}}{\partial Z^{\mu}}\frac{\partial Z^{\nu}}{\partial\hat{Z}^{\beta}}\frac{\partial Z^{\rho}}{\partial\hat{Z}^{\gamma}}\Gamma_{\,\nu\rho}^{\mu}-\frac{\partial Z^{\nu}}{\partial\hat{Z}^{\beta}}\frac{\partial Z^{\rho}}{\partial\hat{Z}^{\gamma}}\frac{\partial^{2}\hat{Z}^{\alpha}}{\partial Z^{\nu}\partial Z^{\rho}}.\label{Gpatching}
\end{equation}
One might expect that we could now identify $H^{0}(PT_{\infty},{\cal G})$
and then proceed to consider the projective vectors of global connections.
Unfortunately there is an issue: the bundle ${\cal G}$ admits no
global sections, as can be seen from, for example, the $_{\,\beta\gamma}^{\alpha}=_{\,\lambda\lambda}^{\lambda}$
component of the patching,
\begin{equation}
\hat{\Gamma}_{\,\lambda\lambda}^{\lambda}=-\lambda^{2}\Gamma_{\,\lambda\lambda}^{\lambda}-4\lambda Q\Gamma_{\, Q\lambda}^{\lambda}-4Q^{2}\Gamma_{\, QQ}^{\lambda}-2\lambda.\label{LLLpatchingG}
\end{equation}
The final term here cannot be (holomorphically) included in any of
the other terms, so there can be no global solutions to (\ref{LLLpatchingG});
$H^{0}(PT_{\infty},{\cal G})=0$.

Inspection of (\ref{Gpatching}) reveals that if one considers only
vectors in the $Z^{A}=(T,Q)$ directions then the patching for the
relevant connection components \emph{does} admit global sections.
Thus it is sensible to decompose the tangent bundle as
\[
T(PT_{\infty})=\mathfrak{h}\oplus\mathfrak{v}
\]
with respect to the fibration $PT_{\infty}\rightarrow\mathbb{CP}^{1}$,
i.e. such that
\[
\beta\in\mathfrak{v}\,\,\mbox{iff}\,\, d\lambda(\beta)=0\,\,.
\]
The general global section of the reduced bundle $\mathcal{G}_{\mathfrak{v}}$
is then
\[
\Gamma_{\, TT}^{T}=\Sigma(T)\qquad\Gamma_{\, TQ}^{Q}=\Gamma_{\, QT}^{Q}=\Xi(T)
\]
\begin{equation}
\Gamma_{\, TT}^{Q}=\Phi_{0}(T)+\Phi_{1}(T)\lambda+\Phi_{2}(T)\lambda^{2}+\Psi(T)Q\,,\label{GeneralGvConnection}
\end{equation}
with all other components $\Gamma_{\, BC}^{A}$ set to zero, and where
$(\Sigma,\Xi,\Phi_{0},\Phi_{1},\Phi_{2},\Psi)$ are six arbitrary
holomorphic functions of $T$.

In the spirit of the spacetime characterisation of the expanded Schrödinger
algebra we will consider, out of all the possibilities in (\ref{GeneralGvConnection}),
the case in which \emph{all} $\Gamma_{\, BC}^{A}=0$. We then define
the subalgebra $S_{\mathfrak{v}}\subset\mathfrak{v}$ to be the algebra
of vertical holomorphic vector fields $\beta$ obeying
\begin{equation}
\mathcal{L}_{\beta}\Gamma_{\, BC}^{A}=\delta_{(B}^{A}\kappa_{C)}\,,\label{Projectivev}
\end{equation}
for $\kappa$ a one-form on $PT_{\infty}$ to be determined in solving
(\ref{Projectivev}). Not unexpectedly, $\kappa$ is always an element
of $H^{0}(PT_{\infty},T^{*}(PT_{\infty}))$, which is populated only
by one-forms $k(T)dT$ for holomorphic $k$ corresponding to the conformal
clock on $M$. The condition (\ref{Projectivev}) sets
\[
a(T)=a_{0}+a_{1}T\quad b(T)=b_{0}+h_{2}T\quad c(T)=c_{0}+c_{1}T
\]
\[
d(T)=d_{0}\quad e(T)=e_{0}+e_{1}T\quad h(T)=h_{0}+h_{1}T+h_{2}T^{2}
\]
for eleven constants $(a_{0},a_{1},b_{0},c_{0},c_{1},d_{0},e_{0},e_{1},h_{0},h_{1},h_{2})$.

To get the twistorial analogue $\tilde{S}$ of the expanded Schrödinger
algebra we must then reintroduce the $\partial_{\lambda}$ parts of
the vectors. This is done by taking the closure under Lie bracket
of
\[
S_{\mathfrak{v}}\oplus\left\{ \left(f(T)+g(T)\lambda+\frac{1}{2}\lambda^{2}d_{0}\right)\frac{\partial}{\partial\lambda}\right\} ,
\]
which fixes
\[
f(T)=f_{0}\qquad g(T)=g_{0}
\]
for two further constants $(f_{0},g_{0})$. The thirteen-dimensional
Lie algebras $\tilde{S}$ on $PT_{\infty}$ and $\widetilde{\mathfrak{sch}}(3)$
on $M$ are then in one-to-one correspondence, a subcorrespondence
of that in Theorem 3.

\subsection{The CGA on $PT_{\infty}$}

In \cite{Lukierski} the authors discuss a particular non-relativistic
limit of the conformal algebra, in which one sends $c\rightarrow\infty$
but scales each generator by an appropriate fact of $c$ such that
the leading term survives. The number of generators is therefore unchanged,
and the resulting fifteen-dimensional algebra is known as the CGA
(conformal Galilean algebra).

We can realise $PT_{\infty}$ as the $c\rightarrow\infty$ limit of
the twistor space $PT_{c}$ associated to Minkowski space \cite{DG15},
and so we can take a limit of the (fifteen) holomorphic vector fields
on $PT_{c}$ in the CGA style to give a representation of the CGA
on $PT_{\infty}$. The conformal Killing vectors on Minkowski space
$M_{c}$ and their resulting limits on $PT_{\infty}$ are shown in
the following table.

\noindent \begin{center}
\noun{}%
\begin{tabular}{|c|c|c|}
\hline 
 & {\small{Vector on $M_{c}$}} & {\small{Limit on $PT_{\infty}$}}\tabularnewline
\hline 
\hline 
{\small{Translations}} & \noun{$\frac{\partial}{\partial t}$} & \noun{$\frac{\partial}{\partial T}$}\tabularnewline
\hline 
 & \noun{$\frac{\partial}{\partial x}$} & \noun{$(\lambda^{2}-1)\frac{\partial}{\partial Q}$}\tabularnewline
\hline 
 & \noun{$\frac{\partial}{\partial y}$} & \noun{$-i(\lambda^{2}+1)\frac{\partial}{\partial Q}$}\tabularnewline
\hline 
 & \noun{$\frac{\partial}{\partial z}$} & \noun{$-2\lambda\frac{\partial}{\partial Q}$}\tabularnewline
\hline 
{\small{Dilation}} & \noun{$t\frac{\partial}{\partial t}+x\frac{\partial}{\partial x}+y\frac{\partial}{\partial y}+z\frac{\partial}{\partial z}$} & \noun{$T\frac{\partial}{\partial T}+Q\frac{\partial}{\partial Q}$}\tabularnewline
\hline 
{\small{Rotations}} & \noun{$x\frac{\partial}{\partial y}-y\frac{\partial}{\partial x}$} & \noun{$iQ\frac{\partial}{\partial Q}+i\lambda\frac{\partial}{\partial\lambda}$}\tabularnewline
\hline 
 & \noun{$y\frac{\partial}{\partial z}-z\frac{\partial}{\partial y}$} & \noun{$-i\lambda Q\frac{\partial}{\partial Q}-\frac{i}{2}(\lambda^{2}-1)\frac{\partial}{\partial\lambda}$}\tabularnewline
\hline 
 & \noun{$z\frac{\partial}{\partial x}-x\frac{\partial}{\partial z}$} & \noun{$-\lambda Q\frac{\partial}{\partial Q}-\frac{1}{2}(1+\lambda^{2})\frac{\partial}{\partial\lambda}$}\tabularnewline
\hline 
{\small{Boosts}} & \noun{$t\frac{\partial}{\partial x}+\frac{x}{c^{2}}\frac{\partial}{\partial t}$} & \noun{$(\lambda^{2}-1)T\frac{\partial}{\partial Q}$}\tabularnewline
\hline 
 & \noun{$t\frac{\partial}{\partial y}+\frac{y}{c^{2}}\frac{\partial}{\partial t}$} & \noun{$-i(1+\lambda^{2})T\frac{\partial}{\partial Q}$}\tabularnewline
\hline 
 & \noun{$t\frac{\partial}{\partial z}+\frac{z}{c^{2}}\frac{\partial}{\partial t}$} & \noun{$-2\lambda T\frac{\partial}{\partial Q}$}\tabularnewline
\hline 
{\small{Special}} & \noun{$-2t\left(x\cdot\partial\right)-\left(x\cdot x\right)\frac{1}{c^{2}}\frac{\partial}{\partial t}$} & \noun{$-T^{2}\frac{\partial}{\partial T}-2TQ\frac{\partial}{\partial Q}$}\tabularnewline
\hline 
 & $\frac{2}{c^{2}}x\left(x\cdot\partial\right)-\frac{1}{c^{2}}\left(x\cdot x\right)\frac{\partial}{\partial x}$ & $(\lambda^{2}-1)T^{2}\frac{\partial}{\partial Q}$\tabularnewline
\hline 
 & $\frac{2}{c^{2}}y\left(x\cdot\partial\right)-\frac{1}{c^{2}}\left(x\cdot x\right)\frac{\partial}{\partial y}$ & $-i(\lambda^{2}+1)T^{2}\frac{\partial}{\partial Q}$\tabularnewline
\hline 
 & $\frac{2}{c^{2}}z\left(x\cdot\partial\right)-\frac{1}{c^{2}}\left(x\cdot x\right)\frac{\partial}{\partial z}$ & $-2\lambda T^{2}\frac{\partial}{\partial Q}$\tabularnewline
\hline 
\end{tabular}\noun{ }
\par\end{center}

The CGA on $PT_{\infty}$ is a finite-dimensional subalgebra of $H^{0}(PT_{\infty},T(PT_{\infty}))$,
giving us another subcorrespondence of that in Theorem 3.

\selectlanguage{british}%

\subsection*{Acknowledgements}

\selectlanguage{english}%
I would like to thank my PhD supervisor Maciej Dunajski who provided
many useful insights over the course of many discussions. Furthermore
I would like to thank Christian Duval for helpful conversations. I
am supported by an STFC studentship.

\end{document}